\documentclass[12pt]{article} 

\usepackage{geometry} 
\usepackage{graphicx} 
\graphicspath{C:/Users/Daniel/Documents/SFU/TrajectoryPaper/}

\usepackage[parfill]{parskip} 
\usepackage{subfig} 
\usepackage{amsmath}
\usepackage{mathtools} 
\usepackage{siunitx}
\usepackage{gensymb}
\usepackage{float}
\usepackage[tableposition=top]{caption}

\usepackage{hanging}
\usepackage{setspace}
\usepackage{authblk}
\title{Rao-Blackwellizing Field Goal Percentage}
\author[1]{Daniel Daly-Grafstein}
\author[1]{Luke Bornn}

\affil[1]{{\normalfont Simon Fraser University}}

\date{October 28, 2018}                 
\begin{document}
\maketitle

\textbf{Abstract} \, Shooting skill in the NBA is typically measured by field goal percentage (FG\%) - the number of makes out of the total number of shots. Even more advanced metrics like true shooting percentage are calculated by counting each player's 2-point, 3-point, and free throw makes and misses, ignoring the spatiotemporal data now available (Kubatko et al. 2007). In this paper we aim to better characterize player shooting skill by introducing a new estimator based on post-shot release shot-make probabilities. Via the Rao-Blackwell theorem, we propose a shot-make probability model that conditions probability estimates on shot trajectory information, thereby reducing the variance of the new estimator relative to standard FG\%. We obtain shooting information by using optical tracking data to estimate three factors for each shot: entry angle, shot depth, and left-right accuracy. Next we use these factors to model shot-make probabilities for all shots in the 2014-15 season, and use these probabilities to produce a Rao-Blackwellized FG\% estimator (RB-FG\%) for each player. We demonstrate that RB-FG\% is better than raw FG\% at predicting 3-point shooting and true-shooting percentages. Overall, we find that conditioning shot-make probabilities on spatial trajectory information stabilizes inference of FG\%, creating the potential to estimate shooting statistics earlier in a season than was previously possible. \par

\section{Introduction} 

Field goal percentage is a common measure of shooting skill and efficiency in the National Basketball Association (NBA), and general shooting prowess is often defined for players by their overall FG\%. It can be used in its raw form, or as a component of more advanced metrics like true-shooting percentage (TS\%) or effective field goal percentage (eFG\%). Shooting percentages play a large role in influencing both fan and coaching evaluation of players, and are often used to predict future player performance when making decisions regarding free agency or draft selection.   

Predicting a player's FG\% given past shooting is a difficult task. Shooting percentages are highly variable, especially on longer shots like 3-point attempts. For example, it takes roughly 750 3-point attempts before a player's shooting percentage stabilizes, where over half of the variation in their 3-point percentage (3P\%) is explained by shooting skill, rather than noise (Blackport 2014). Additionally, 3P\% has been shown to be an unreliable metric in terms of its ability to discriminate between players and its stability from one season to the next (Franks et al. 2016). As the proportion of shot attempts taken as 3-pointers increases, with total attempts having risen nearly 50\% over the last 8 years (Young 2017), overall FG\% becomes more variable and less stable. 

Part of the large variation in shooting percentages is likely due to the many contextual factors that contribute to the probability of a shot make. Improvements to FG\% prediction have been made by including some of these covariates in shot-make prediction models (Cen et al. 2015, Piette et al. 2010). However, because of the small differences that separate the true shooting skill of players in the NBA, chance variation may also contribute significantly to the variation and instability of FG\%. Optical tracking data of shot trajectories can potentially reduce noise in shooting metrics by allowing us to differentiate shots that rim out, air balls, and (unintentional) banks, giving us more information about players' shooting skill with fewer shots. This idea has been demonstrated recently during practice shooting sessions, where FG\% augmented by precise shot factor information gathered during these sessions improved the prediction of future shooting (Marty and Lucey 2017, Marty 2018). Accurate estimates of shot factors using live-game optical tracking data may allow for a similar improvement in the prediction of in-game shooting metrics. \par

In this paper, we seek to reduce the variation in predicting player FG\% using NBA optical tracking data. We begin the paper by introducing a new estimator for FG\%, RB-FG\%, based on aggregating shot-make probabilities. Estimation of shot-make probabilities is then split into two main parts. First, using spatio-temporal information provided by the tracking data, we model shot trajectories in order to estimate the depth, left-right distance, and entry angle of balls entering the basket. Next, we use a regression model to estimate the probability of each shot going in. We define the average of these estimated probabilties, RB-FG\%, as our new estimator of FG\% for each player. Finally, we compare the predictive ability of the RB-FG\% estimator to its raw counterpart that does not utilize trajectory information. \par

\section{The Rao-Blackwellized Estimator} 

In this section we introduce our new estimator for FG\% based on shot-make probabilities. When trying to predict a player's future FG\% using their past FG\%, each shot $X_i$ is treated as Bernoulli random variable with probability of success $\theta$, where $\theta$ is a measure of the player's true FG\%. However, shot trajectories provided by optical tracking data gives us more information for each shot than simply whether it is a make or a miss. Incorporating this information into a shot model may allow us to reduce the variance involved in estimating and predicting shooting skill. Therefore, we can define an alternative model where the probability of a shot-make varies depending on its trajectory, and shots are modeled as Beta-Bernoulli random variables $X_i \sim Bern(p_i)$ with $p_i \sim Beta(\theta v, (1-\theta) v)$, where again $\theta$ is the true FG\% of a player, defining their corresponding Beta distribution of shot-make probabilties. Each player's shooting ability is now modeled by a Beta distribution, and the probability of a shot going in follows a Bernoulli distribution indexed by $p_i$, where $p_i$ is a draw from that player's Beta distribution.

As shown below, inference under the model in which shots are treated as Bernoulli random variables and inference under the expected Beta-Binomial of our new model is the same. Let $\Pi(X_i | \theta,v)$ be the likelihood of the expected Beta-Binomial distribution, i.e. the likelihood of the Beta-Binomial distribution if you mariginalize out the $p_i$'s,  and let $B(\cdot)$ be the beta function. \par
\vspace{-0.5cm}

\begin{align}
\Pi(X_i | \theta,v) &\propto \int_{0}^{1} \frac{p_i^{X_i+\theta v-1}(1-p_i)^{(1-\theta)v-X_i}}{B(\theta v, (1-\theta) v)} dp_i \notag\\[10pt]
&\propto \frac{B(X_i + \theta v, 1-X_i + (1-\theta)v}{B(\theta v, (1-\theta)v)}\notag\\
&\propto \theta^{X_i}(1-\theta)^{(1-X_i)} \notag
\end{align}

\vspace{-0.5cm}

Therefore, inference for $\theta$ is the same under the Bernoulli and expected Beta-Binomial distributions. Furthermore, suppose we obtain $X_i$ (make or miss) and $p_i$ (the probability that shot $i$ will go in). Let $\Pi(X_i,p_i|\theta,v)$ be the joint distribution of $X_i$ and $p_i$. It follows that:
\vspace{-0.9cm}

$$\Pi(X_i,p_i|\theta,v) = \Pi(X_i|p_i)\Pi(p_i|\theta,v)$$
\vspace{-1.2cm}

where $\Pi(X_i|p_i)$ and $\Pi(p_i|\theta,v)$ are the Bernoulli and Beta distributions, respectively.  Consequently we have that given $p_i$, $X_i$ is independent of $\theta$. Thus $p_i$ is sufficient for $\theta$. Now let $\hat{\theta}$ be the raw FG\% estimate and $\hat{\theta}_{RB}$ be the RB-FG\% estimate. We have:
\vspace{-0.8cm}

\begin{align}
\hspace{.25cm} \hat{\theta} &= \frac{1}{N} \sum_{i=1}^{N} X_i \hspace{0.5cm} \text{(FG\%)} \notag\\
\hat{\theta}_{RB} = E\left(\hat{\theta}\hspace{0.01cm} |p_1,..., p_N\right) &=\frac{1}{N} \sum_{i=1}^{N} p_i \hspace{0.5cm} \text{(RB-FG\%)} \notag\\
\notag
\end{align}

\vspace{-1.3cm}
Thus the RB-FG\% is simply the conditional expectation of raw FG\% given these shot-make probabilities $p_i$. Because under the Beta-Binomial model $p_i$ is sufficient for $\theta$, by the Rao-Blackwell Theorem we have:
\vspace{-0.6cm}

$$ MSE\left(\hat{\theta}_{RB}\right) \leq MSE\left( \hat{\theta}\right) $$ \par

\vspace{-0.2cm}
Unfortunately, we are unable to know the true probability that a shot will go in. Therefore, as decribed below, we will use estimates of shot-make probabilities based on shot trajectory information to obtain an estimate of RB-FG\%. Using an estimate of $\hat{\theta}_{RB}$ means that the inequality above does not necessarily hold. However, as we will see in section 4, our estimates of shot-make probabilites are accurate and precise enough that this estimate of $\hat{\theta}_{RB}$ still leads to a decrease in variance and prediction error relative to raw FG\%. For simplicity, moving forward we will refer to the statistic based on estimated shot-make probabilities as $\hat{\theta}_{RB}$.

\section{Estimating Shot-Make Probabilities}

\subsection{Measuring Shot Factors}

In order to estimate shot-make probabilities, we first measure three shot factors based on how each shot entered the basket - left-right accuracy, depth, and entry angle - following the procedure of Marty and Lucey (2017). We define left-right accuracy as the deviation of the ball from the centre of the hoop as the ball crosses the plane of the basket (Figure 1a). Shot depth is defined as the distance of the ball from a tangent line through the front of the hoop as the ball crosses the plane of the basket (Figure 1a), with the front of the hoop adjusted to be from the perspective of the shooter. We specify the adjusted front of the rim as depth 0, so a shot crossing the basket plane at the center of the hoop has a depth of 9 inches.  Finally, the entry angle is defined as the angle between the plane of the hoop and a tangent line through the ball as it is entering the basket (Figure 1b). See Marty and Lucey (2017) for further detail regarding these measurements. 

\begin{figure}[!b]
  \centering
   \subfloat[]{\includegraphics[width=9cm,height=4.5cm]{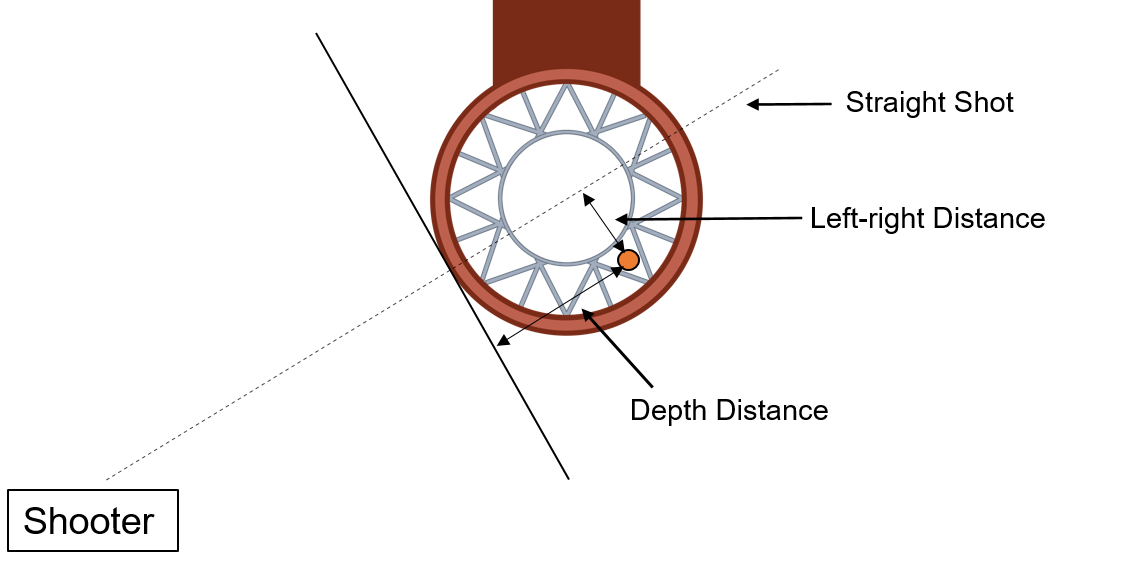}\label{fig:f1}}
  \subfloat[]{\hspace{-0.5cm}\includegraphics[width=8cm,height=6cm,trim={6cm 2cm 6cm 2cm},clip]{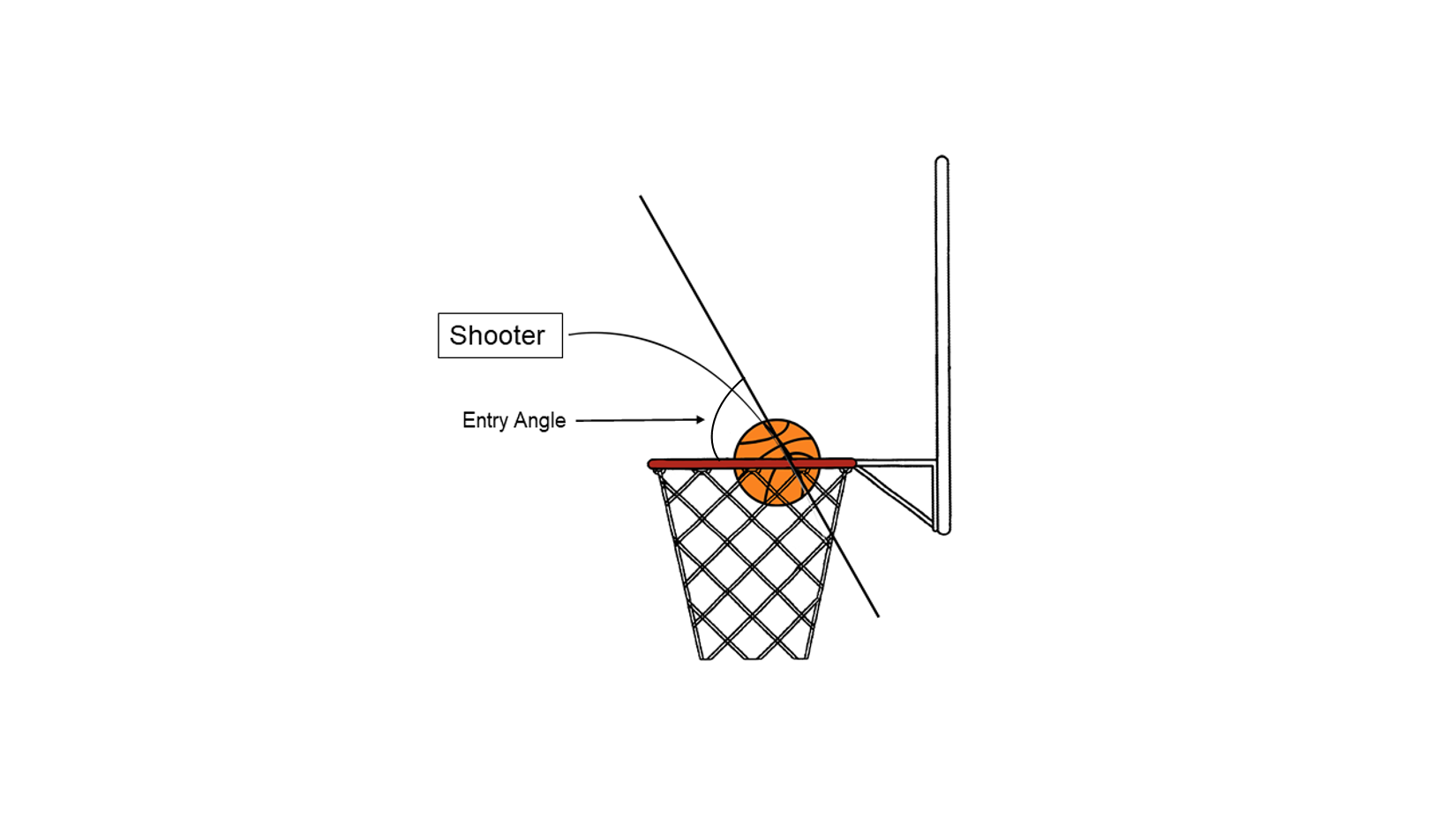}\label{fig:f2}}
  \caption{Shot factors at the plane of the hoop. Figure (a) denotes the left-right and depth factors, Figure (b) denotes the entry angle factor.}
\end{figure}

To obtain these shot factor estimates, we use shot trajectory information provided by the SportVu optical tracking data from STATS LLC. The data provides measurements of the X and Y coordinates for all 10 players and X,Y, and Z coordinates of the ball 25 times per second. Our dataset consists of 1212 games from the 2014-15 NBA regular season and 1206 games from the 2015-16 regular season. We first restrict our analysis to 3-point shots as these shots have the most trajectory information and we can assume all shooters are attempting to hit the centre of the basket (no shot attempts purposely off the backboard). In total our dataset consists of trajectory information for 47,631 3-point shots from the 2014-15 season and 49,876 3-point shots from the 2015-16 season. \par

Although the optical tracking data gives X,Y, and Z coordinates of the ball at the basket, the location data is noisy, especially in measuring the height of the ball. To obtain a better estimate of the position of the ball near the basket we model a quadratic best fit line through the trajectory data given by the tracking database. If $ Z_{i}$ is the height of shot $i$, and $x_i$ and $y_i$ are the X,Y coordinates of the shot in the tracking data, we use a quadratic polynomial to model the height, and estimate the coefficients by a least-squares regression:

\begin{equation}
E(Z_i) = \beta_0 + \beta_1x_i + \beta_2y_i + \beta_3x_i^2+ \beta_4y_i^2 +\beta_5x_iy_i 
\end{equation}

We use the point where the model specifies the ball crosses 10 feet in height as the estimated X,Y location of the ball at the basket, and use this location to calculate the shot's depth, left-right accuracy, and entry angle. \par

We compare the above model with a second model in which we try to leverage pre-existing knowledge of shot trajectories. We know each shot starts at the player's location at the time of release (player location is less noisy than ball location in the tracking database) and ends around the basket. Therefore, we can improve estimation by biasing the start and end points of our modeled trajectories to incorporate this prior knowledge. To accomplish this we introduce a Bayesian regression model using pseudo-data to establish priors that reflect this knowledge. This is an informal empirical Bayes method where instead of using data to estimate the priors, we use prior knowledge of how the data should look. Given the quadratic model (1) for each shot, we can specify a Bayesian regression model with a conjugate Normal prior for $\beta$ of the form $\rho(\beta|\sigma^2,z,X) \sim N(u_{0}, \, \sigma^2\Lambda_{0}^{-1})$. This results in a conjugate inverse gamma prior for $\sigma^2$ written as $\rho(\sigma^2|z,X) \sim IG(a_{0},b_{0})$. We can then update our mean and precision parameters as:

$$u_{n}=\left(X^TX+\Lambda_0\right)^{-1}\left(\Lambda_0u_0+X^TX\hat{\beta}\right) ,    \hspace{0.4cm} \Lambda_n=\left(X^{T}X+\Lambda_0\right)$$

where $u_n$ is the posterior mean of $\beta$, and $\Lambda_n$ is the posterior precision matrix for $\beta$. We update the parameters twice, once using pseudo-data reflecting our prior knowledge of where shots start and finish, and a second time using the shot trajectory data from the optical tracking data. We specify 4 pseudo-data points, 2 at the start of the shot set at the X,Y coordinates of the player when the shot is released and at a height of 7 feet, and 2 set at the centre of the hoop and at 10 feet in height. After two Bayesian learning updates we take the posterior mean of $\beta$, $u_2$, and use it as the estimate for the coefficients in the quadratic polynomial model (1). \par

\begin{figure}[t]
  \centering
  \includegraphics[width=15cm]{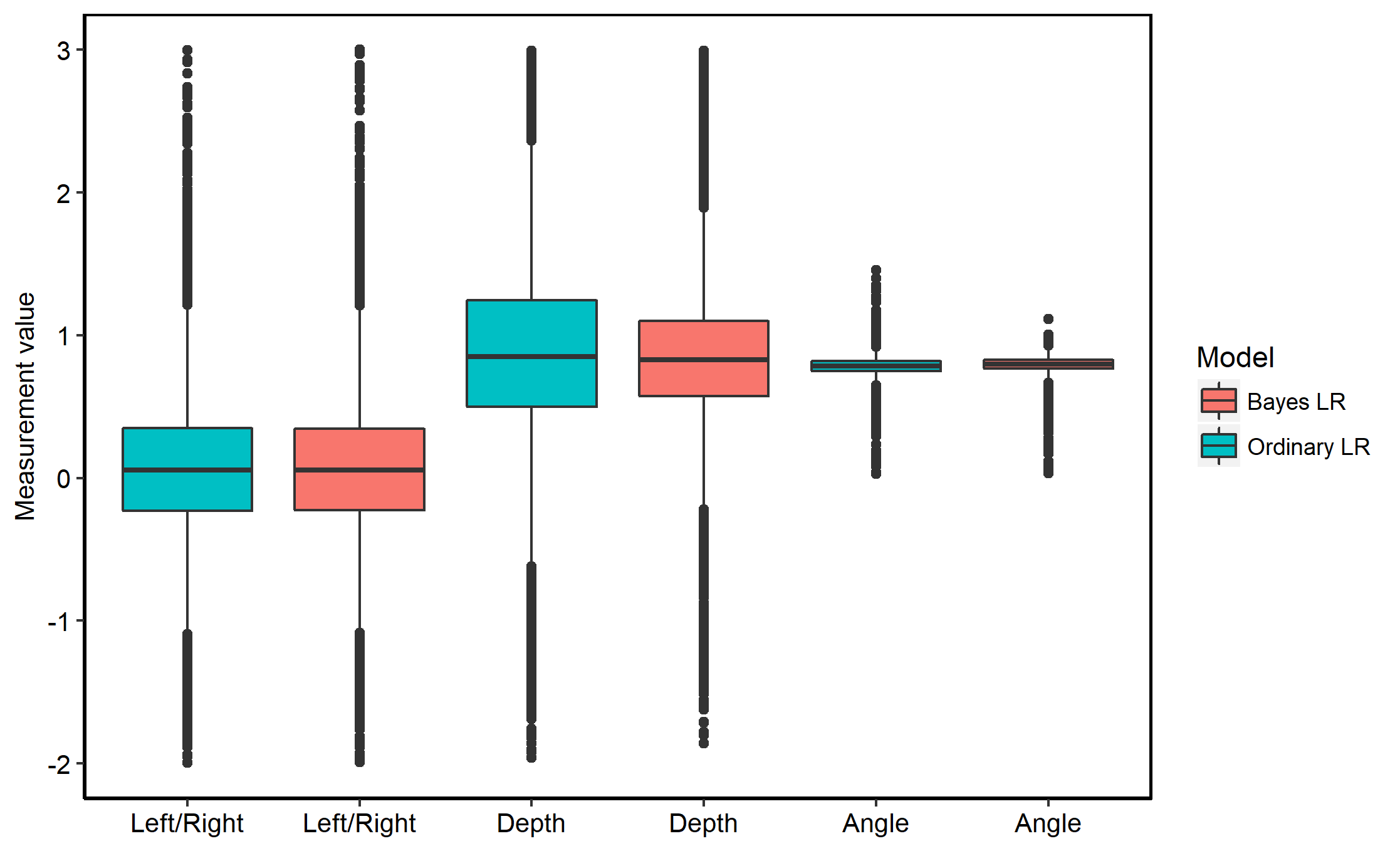}
  \caption{Estimated shot factor measurements under the ordinary and Bayesian regression models. Left/right and depth measurements are given as distance in feet in relation to the centre of the basket, where the depth of the centre of the basket is 0.75 feet. Entry angle measurements are given in radians to allow for the values to be on the same axis.}
  \label{fig:shotfactors}
\end{figure}

We then use (1) to compute the 3 shot factors for each shot using both the ordinary linear regression (OLR) and Bayesian regression approaches. Comparing the two models, we find both predict shots to have a mean depth value of 11", a mean left/right value of 0", and a mean entry angle around \ang{45}. As in Marty and Lucey (2017) we find shots entering the basket at 11" in depth, 2" deeper than the centre of the basket, and 0" in left-right accuracy are made with the highest percentage. However, we find shot depths are evenly distributed around 11", in contrast to the findings of Marty and Lucey (2017) who found that shooters have a mean shot depth value of 9", at the centre of the hoop. The variance in left/right distance and entry angle between the two models is similar, however the variance in shot depth is much larger in the OLR compared to the Bayesian regression model (Figure 2). Overall, variances in shot factors under the Bayesian model match the variances of the precise shot factor measurements of Marty and Lucey (2017) more closely than the OLR model. Furthermore, we will see later that when we model shot probabilities the Bayesian model produces a lower misclassification rate and log loss than the OLR model. Moving forward, we decide to use shot factors calculated via the Bayesian regression model.

We next compare the precision of our estimated shot factors to those measured by the Noah Shooting System - a dedicated hardware install found in practice facilities that provides shooting information not available in live games. Marty and Lucey (2017) were able to use the Noah system to define a Guaranteed Make Zone (GMZ) of over 90\% based on these shot factors. Their GMZ is marked by shots with an entry angle of \ang{45}, a left-right accuracy between -2" and 2", and a depth between 7" and 14". Using our estimated shot factors, we found shots in this GMZ are made only 85.2\% of the time. This suggests that despite the Bayesian model, our shot factor estimates are still less precise than those gathered by the Noah system. \par

\subsection{Modeling Shot-Make Probabilities} 

In this section we train a shot-make probability model using 3-point shots from the 2014-15 season. To obtain shot-make probabilities for each shot, we use the estimated shot factors described previously as covariates in a logistic regression:

\begin{equation} 
P(S_i=1)=\sigma \Bigg( \Bigg.
\begin{array}{c l}	
 &\beta_0 + \beta_1 \hat{D}_i + \beta_2 \hat{LR}_i + \beta_3 \hat{A}_i + \beta_4 \hat{D}_i^{2} + \beta_5 \hat{LR}_i^{2}\\
 &+\beta_6 \hat{A}_i^{2} +\beta_7 \hat{D}_i*\hat{LR}_i + \beta_8 \hat{D}_i*\hat{A}_i + \beta_9 \hat{LR}_i*\hat{A}_i 
\end{array}\Bigg. \Bigg)
\end{equation}

where $S_i$ is an indicator function equal to 1 when a shot goes in and 0 went it misses, $i$ indexes all 3-point shots from the 2014-15 season (N=47,631), $\sigma(\text{x}) = \text{exp(x)}/(1+\text{exp(x)})$, and $\hat{D}_i$, $\hat{LR}_i$, and $\hat{A}_i$ are the estimated depth, left-right distance, and entry angle of shot $i$, respectively. We note that the Rao-Blackwell inequality indicates the framework detailed in section 2 holds regardless of the choice of shot probability model, given the model provides reasonable estimates of shot probabilities. \par   

Although our Bayesian regression model biases shot trajectories toward the basket, some trajectories are still quite variable. Modeled trajectories that are too far from the raw data are removed and instead assigned a probability of 1 or 0 for a make or miss, respectively. We use factors from the remaining shots to estimate shot-make probabilities with model (2).  To assess how accurate the model is we perform a tenfold cross validation to obtain the mean misclassification rate, as well as calculate the log loss and Brier score. We repeat this procedure with shot factors estimated from the OLR model, and the results are shown in Table 1. \par

\newcommand\Tstrut{\rule{0pt}{2.6ex}}         
\newcommand\Bstrut{\rule[-0.9ex]{0pt}{0pt}} 

\begin{table}[!b]
\caption{Mean Misclassification Rate, Brier Score, and Log Loss of Model (2)}
\centering
\begin{tabular}{cccc}
  \hline \hline
 & Misclassification Rate & Brier Score & Log Loss \Tstrut\Bstrut\\
  \hline
  Grand Mean & NA & 0.228 & 0.648 \Tstrut\Bstrut\\
  OLR & 0.246 & 0.176 & 0.528  \Tstrut\\ [1ex]
  Bayesian Regression & 0.204 & 0.160 & 0.491  \\ [1ex]
   \hline
\end{tabular}
\begin{flushleft}
Log loss and Brier scores are based on shot-make probability predictions from model (2) for 3-point shots from the 2015-16 NBA season. The covariates are estimated via the Bayesian regression and OLR methods described in Section 3.1, while the Grand Mean is the league-wide 3P\% for the 2014-15 season. The mean misclassification rate is the result of tenfold cross validation.
\end{flushleft}
\end{table}

The covariates estimated via Bayesian regression resulted in misclassification rate 0.204. Therefore, our Bayesian model is able to predict makes/misses correctly about 80\% of the time. This is a higher rate than many shot prediction models that use contextual covariates, like those presented in Cen et al. (2015) which utilize variables such as distance to basket and nearest defender to predict shot-makes with 65\% accuracy. Similar to probabilities based on raw FG\% (Marty and Lucey 2017), predicted shot-make probabilities are highest for shots at 11 inches depth, 0 inches of left-right deviation, and similar for shots with entry angles in the mid-40s. These can be seen in relation to the basket in Figure 3. \par

\begin{figure}[!b]
  \centering
  \subfloat[]{\includegraphics[width=7cm,height=5.5cm]{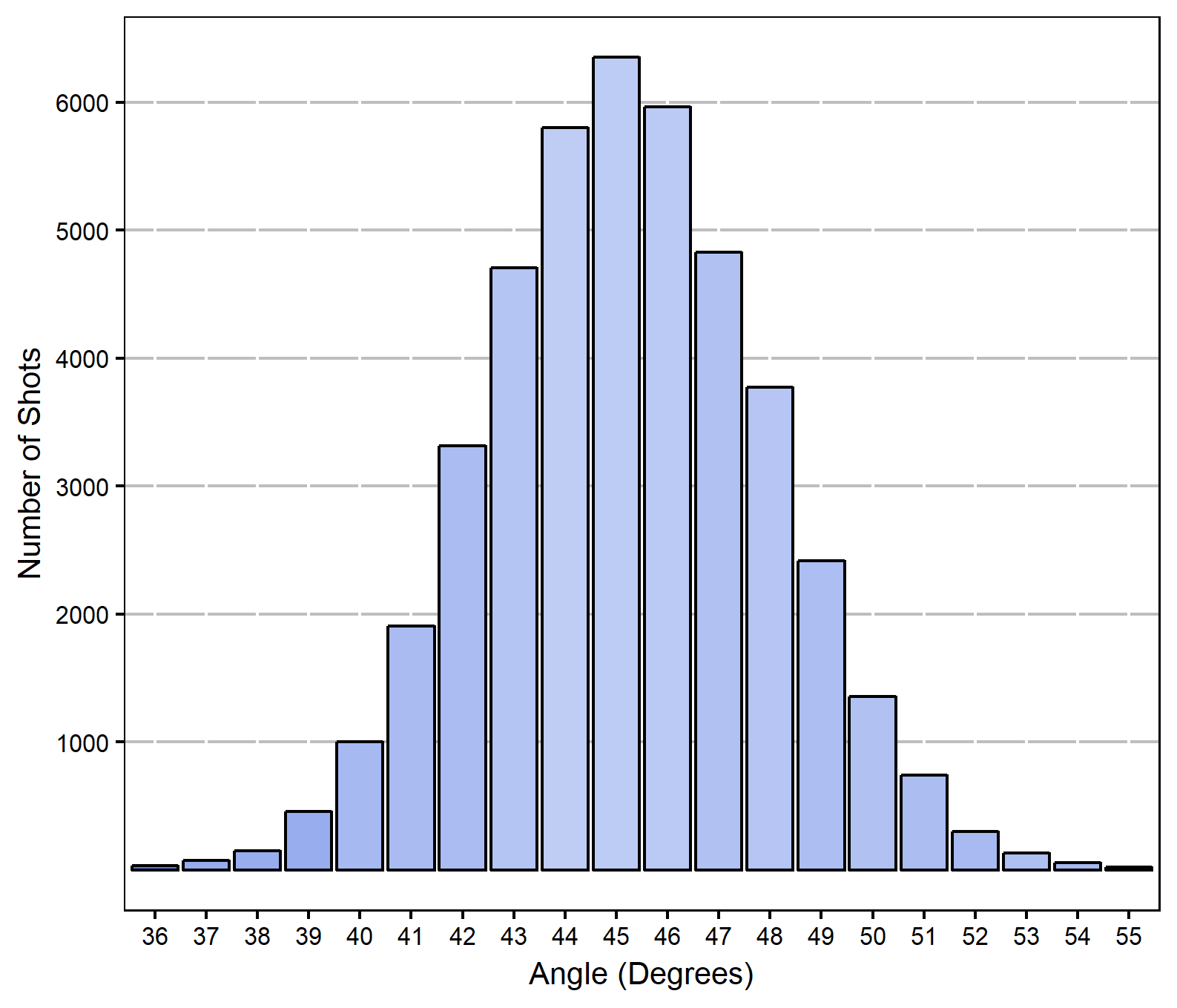}\label{fig:f3}}
  \subfloat[]{\includegraphics[width=7.6cm,height=6.1cm]{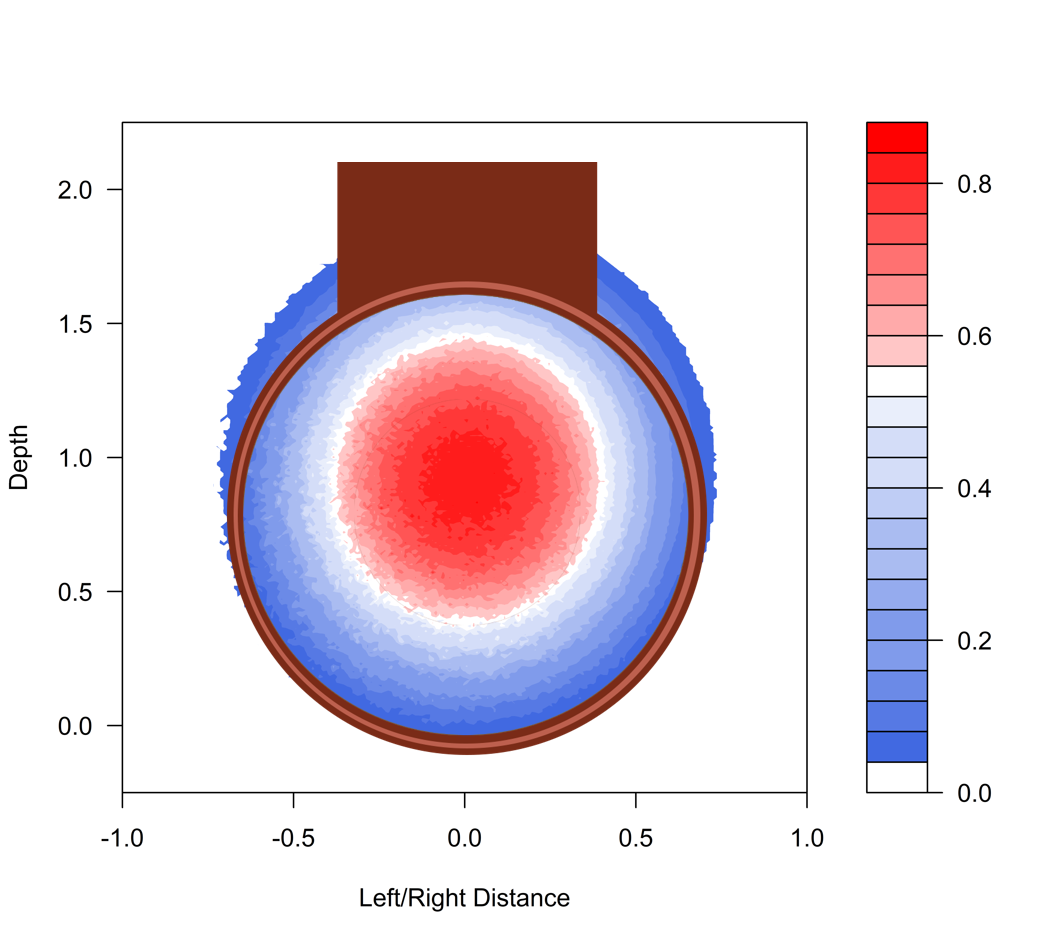}\label{fig:f3}}
  \caption{Figure (a) shows the distribution of mean predicted shot-make probabilities over different shot entry angles. Included are all 3-point shots in the 2014-15 season in which trajectory information is used to train our model (2). Figure (b) shows the distribution of predicted shot-make probabilities over different values of shot depth and left-right accuracy in relation to the basket.}
\end{figure}

\section{Applications of the RB-FG\% Estimator}

\subsection{Predicting Three-Point Field Goal Percentage}

In this section we aim to create a new estimate for player FG\% by aggregating estimated shot-make probabilities given by (2). Without loss of generality, we focus first on 3-point shots for clarity of presentation. We gather shot trajectories for 3-point shots taken from the first half of the 2015-16 NBA season in the SportVu tracking database (N=24855), and predict the probability of each shot going in using model (2) trained by shots taken in the 2014-15 season. The mean of these estimated shot-make probabilities is the RB-FG\% estimate, $\hat{\theta}_{RB}$, for each player's FG\%. We wish to see whether $\hat{\theta}_{RB}$ is better than raw FG\%, $\hat{\theta}$, at predicting a player's future FG\%, $\theta$. We find that when predicting 3-point FG\% in the second half of the 2015-16 season, $\hat{\theta}_{RB}$ outperforms $\hat{\theta}$ in terms of mean absolute error (Table 2). Interestingly, as seen similarly in Brown (2008), $\hat{\theta}$ is quite a poor predictor of future shooting. It performs worse than simply using the league-wide grand mean as a predictor for every player (Table 2). Additionally, as both $\hat{\theta}$ and $\hat{\theta}_{RB}$ are unbiased estimators of $\theta$, the decrease in mean-absolute error of our modeled estimator is predominately due to a reduction in variance (Figure 4). Furthermore, we can see that the uncertainty inherent in our shot factors estimates, and thus our shot-make probability estimates, add very little to the overall variance in $\hat{\theta}_{RB}$. \par

In addition to assessing prediction accuracy, we can also investigate whether the RB-FG\% estimator produces more consistent player rankings than raw FG\%. We calculate $\hat{\theta}$ and $\hat{\theta}_{RB}$ for 3P\% in the first and second half of the 2015-16 season and rank all 260 players in our analysis according to each estimate. The $\hat{\theta}$ and $\hat{\theta}_{RB}$ estimates produce Spearman's rank coefficients of 0.216 and 0.245, respectively. We find, using the tests detailed in Fieller et al. (1957), that although RB-FG\% produces a higher rank correlation than raw FG\%, it is not significantly higher.


\begin{figure*}[!t]
  \centering
  \includegraphics[width=13cm]{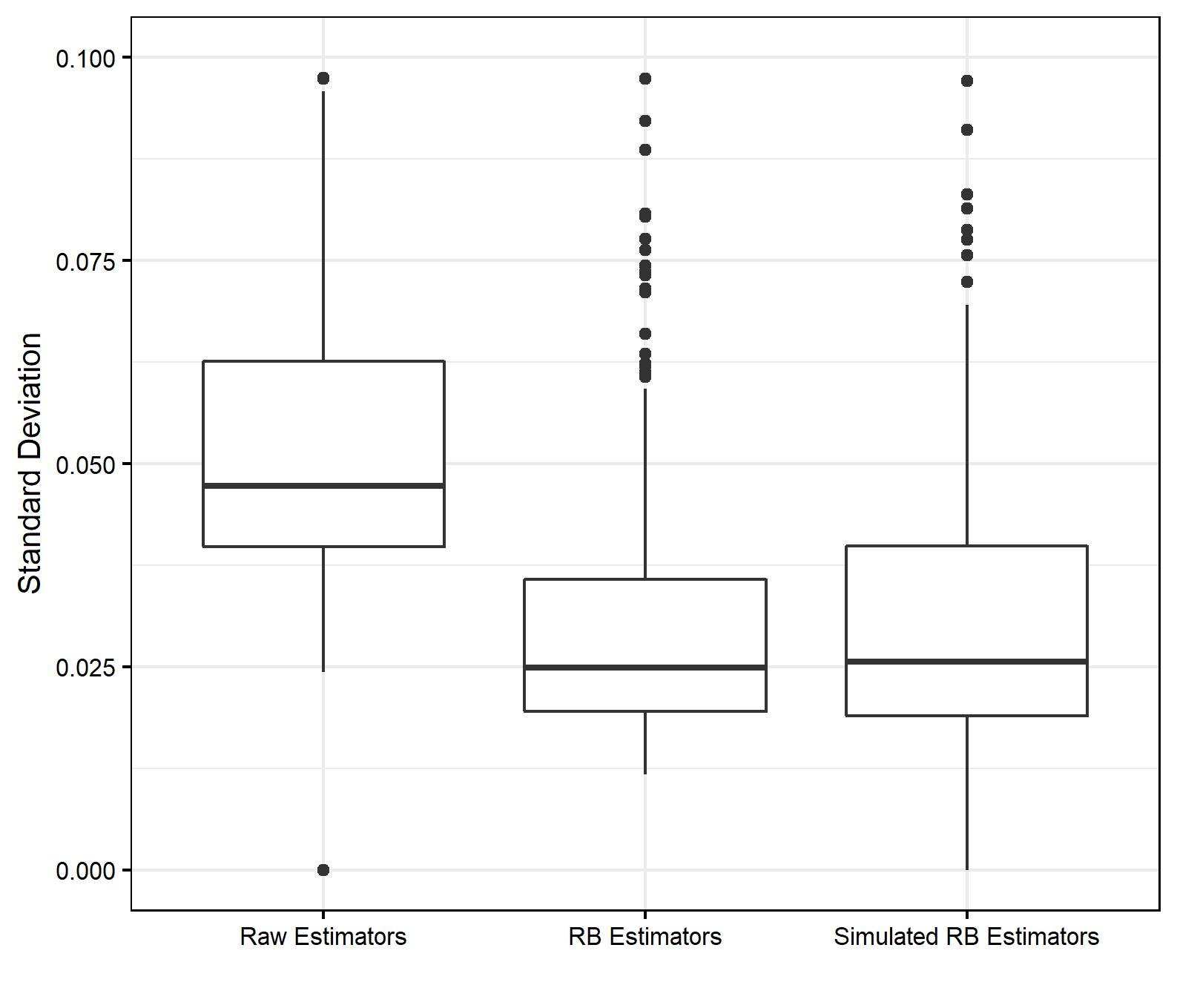}
  \caption{The distribution of standard deviations for the Rao-Blackwellized (RB), raw 3-point FG\%, and simulated RB estimators for 260 players in the first half of the 2015-16 season. The variance of the raw estimator for each player is calculated according to a binomial distribution as $\hat{\theta}(1-\hat{\theta})/n$, where $\hat{\theta}$ is the raw FG\% for each player. The variance of the Rao-Blackwellized estimator is calculated according to the proposed Beta distribution that each player's shot probabilities are drawn from as $(\hat{\theta}\hat{v})/(n(\hat{\theta}+\hat{v})^{2}(\hat{\theta}+\hat{v}+1))$, where $\hat{\theta}$ and $\hat{v}$ are calculated by maximum likelihood using Nelder-Mead optimization (Griffiths 1973, Skellam 1948). The Simulated RB boxplot denotes the standard deviations of the RB estimators taking the uncertainty of shot factor estimates into account. These standard deviations are calculated by first resampling shot factors from the multivariate normal distributions on the parameters in (1), and recalculating shot probabilities via (2). Next, we take a sample of these new shot-make probabilities and calculate the standard deviation of the simulated estimator empirically. We repeat this process 10 times and take the average to get an estimate of the standard deviation of the estimator for each player.}
 \label{fig:Variance}
\end{figure*}

\begin{figure}[!b]
  \centering
  \includegraphics[width=13cm]{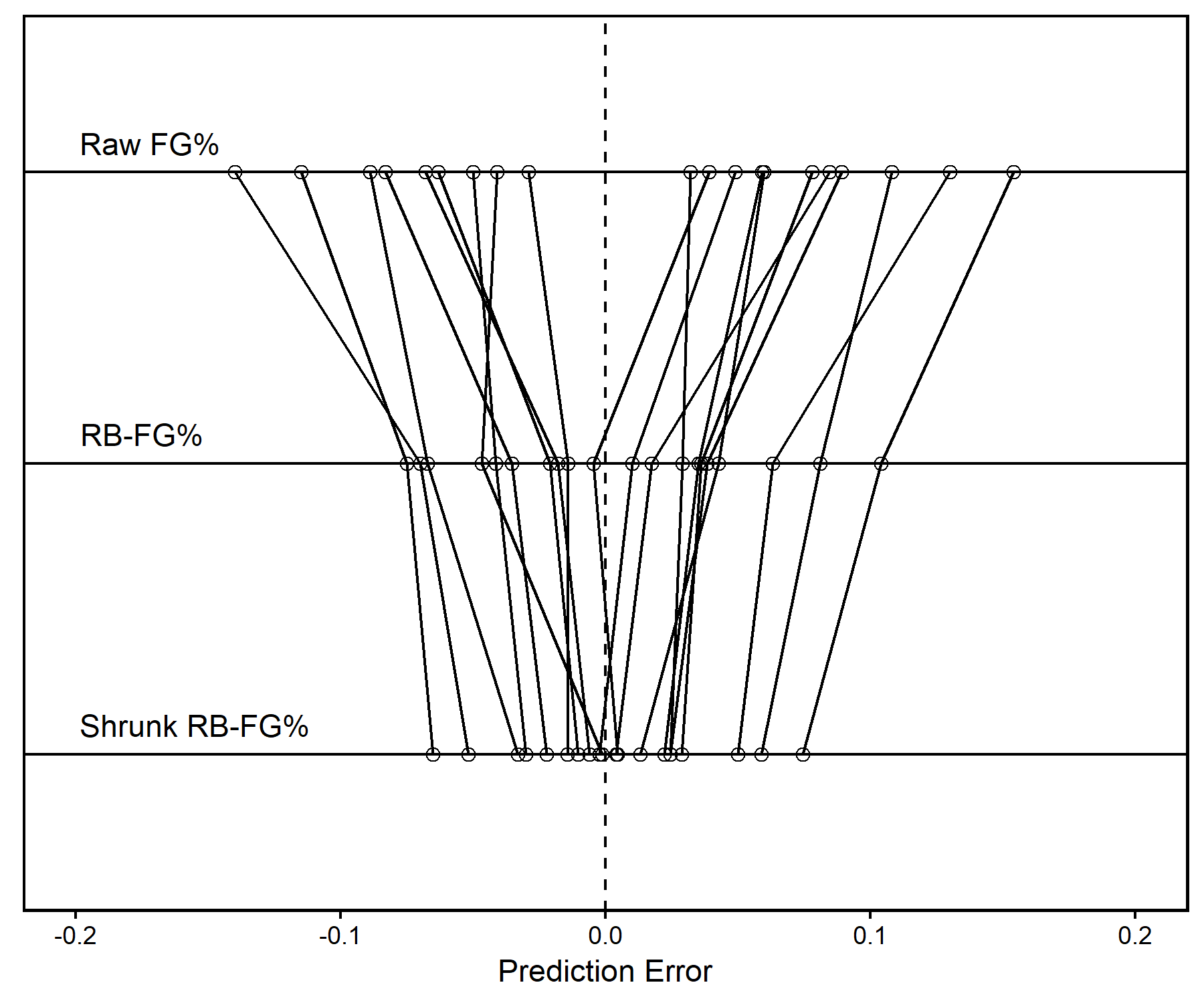}
  \caption{Mean prediction error for the raw, Rao-Blackwellized, and shrunk-Rao-Blackwellized 3-point FG\% estimators of 20 players in the first half of the 2015-16 season. Errors are measured for predicting 3-point FG\% in the second half of 2015-16.}
  \label{fig:Shrinkage}
\end{figure}

Rao-Blackwellizing the estimator for FG\% does reduce variance and improve the prediction accuracy, but these estimators are based on low sample sizes for most players. Players in our dataset take between 3 and 402 three-point attempts in the first half of the 2015-16 season, far fewer than the number needed for 3P\% to stabilize (see section 1). We are able to further reduce the variance of $\hat{\theta}_{RB}$ by introducing a empirical Bayesian shrinkage factor towards a Beta prior, $B(\alpha_0,\beta_0)$ (Casella 1985). We choose the hyperparameters of the Beta prior based on the posterior mean 3P\% in the first half of the 2015-16 season (0.35), and tune $\alpha_0$ in terms of minimizing the mean absolute error of $\hat{\theta}_{RB}$.  We end up applying a prior distribution to each player's first half 3-point shooting of the form $B(3.5,6.5)$, in essence adding 10 league-average shots to $\hat{\theta}$ and $\hat{\theta}_{RB}$. Shrunk-RB estimates are calculated by the expected value of the updated Beta distribution as:

\begin{equation}
\hat{\theta}_{Shrunk{\text -}RB} = \frac{3.5+\hat{\theta}\hat{v}}{3.5 + 6.5 +\hat{\theta}\hat{v}+(1-\hat{\theta})\hat{v}}
\end{equation}

Table 2 shows that the shrunk-RB estimator is a better predictor than the shrunk-raw estimator, and this improvement is illustrated in Figure 5. Hence while Rao-Blackwellizing significantly improves prediction, leveraging knowledge about the distribution of 3P\%'s can further improve the RB-FG\% estimator  (Morris and Effron 1977). \par

In addition to predicting future shooting, we can also use $\hat{\theta}_{RB}$ to estimate players' 3P\% with less data than when using $\hat{\theta}$. The root-mean-square error (RMSE) of both estimators for inferring end-of-season 3P\% is presented in Figure 6. RB-FG\% has a lower RMSE than FG\% when calculated using less than 30\% of games, and the biggest improvements occur with low sample sizes. Some bias is introduced by RB-FG\% as shot probabilities are modeled in (2) using the entire set of 3-point shots, while estimates are calculated seperately for each player. We only used a single set of priors to estimate shot factors in our Bayesian regression, but each player should have their own set of priors due to differences in height and shooting style. However, specifying individual priors and creating separate shot trajectory models for each player is difficult because most players take too few shots to obtain accurate parameter estimates. Additionally, the reduction in variance outweighs the small level of bias introduced by $\hat{\theta}_{RB}$ (Figure 6). Because we are comparing these estimators to raw FG\% on the full sample, the raw estimator becomes better if we calculate RMSE using more than 40\% of games. However, even full-season shooting numbers are highly variable and based on low sample sizes for most players. Thus RB-FG\% is a better overall estimate on any size of data, but for small sample sizes it is a better estimate of end-of-season FG\% than FG\% itself. \par
\vspace{1cm}

\begin{figure*}[!b]
  \centering
  \hspace*{0.5cm}\includegraphics[width=15cm]{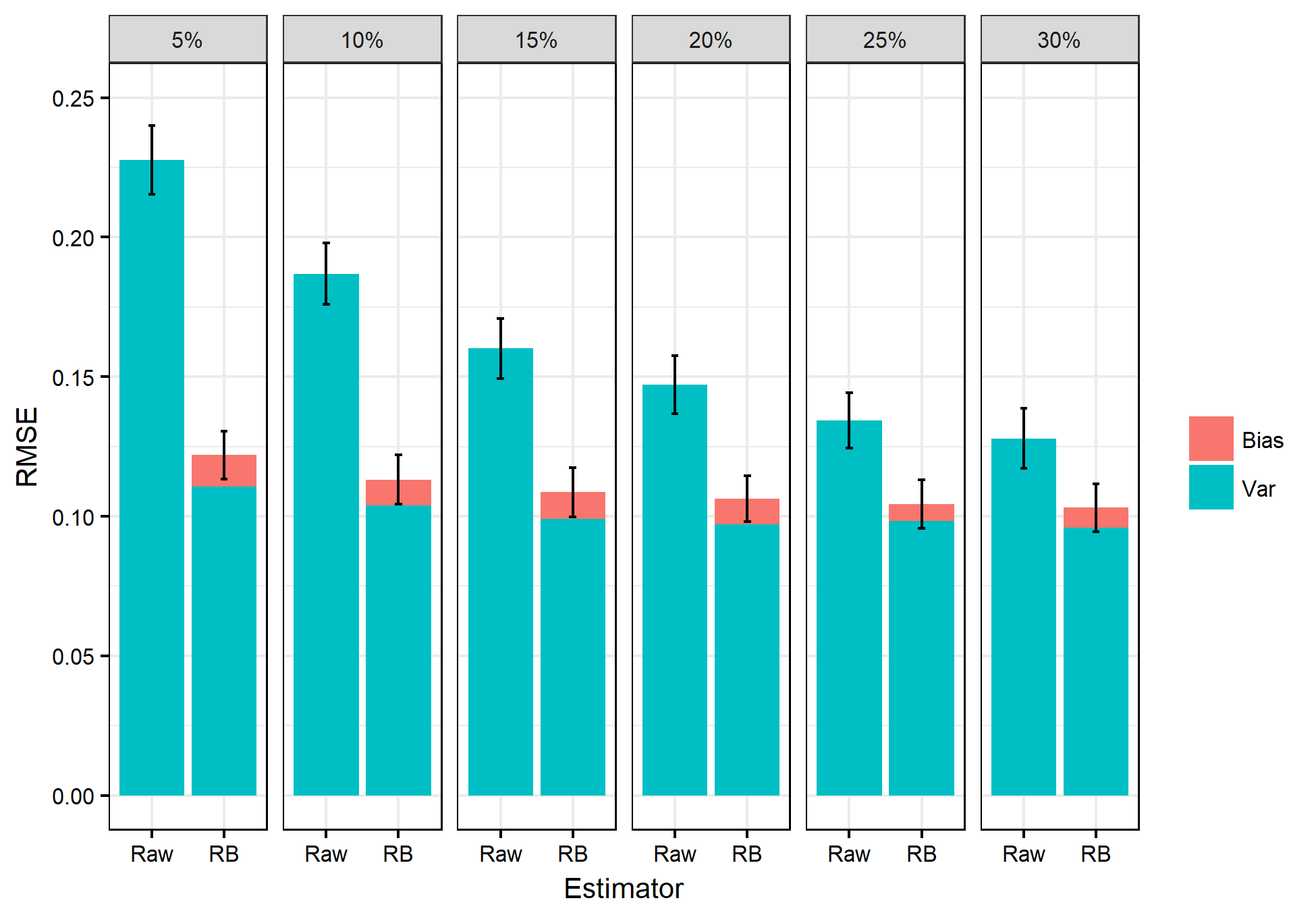}
  \caption{The RMSE of $\hat{\theta}$ and $\hat{\theta}_{RB}$ estimating players' true 3-point FG\% for the 2014-15 NBA season. These estimators are calculated using shots from a subset of games and compared to each player's 3-point FG\% at the end of the season. RMSE is calculated separately for each sub-box using 5\%, 10\%, 15\%, 20\%, 25\%, and 30\% of the games from the 2014-15 season.}
  \label{fig:Shrinkage}
\end{figure*}

\subsection{Predicting True Shooting Percentage}

Although we've focused on three-point shots, we are able to Rao-Blackwellize any shooting statistic provided we have enough trajectory information to accurately estimate shot factors. We now expand our selection of shots and try to improve predictions of TS\% using our shot factor and shot probability models. We repeat the procedure described in section 3 to estimate shot factors for all two-point shots and free throws in the 2014-15 season, and use these to create separate Rao-Blackwellized two-point FG\% and free throw percentage (FT\%) estimates. As before, shots that do not have enough location data or resulted in trajectory predictions very far from the raw data are not included in training or prediction datasets. In total, shot-make probabilities are estimated for 21,153 out of 24,832 free throws and 21,890 out of 73,925 two-point shots, with remaining probabilities assigned as 1 or 0 for a shot make and miss, respectively. The new RB estimators are again used to predict two-point FG\%, FT\% and TS\% in the second half of the 2015-16 season. As with 3P\%, we find the shrunk Rao-Blackwellized estimator for TS\% results in the lowest mean absolute error (Table 2). \par

Rao-Blackwellizing 2-point shots results in only a modest decrease in MAE compared to the shrunk raw estimator. This may be because we are only able to estimate shot-make probabilities for a small fraction of two-point shots using the optical tracking database. Many 2-point shots are taken close to the basket or intended as bank-shots, resulting in insufficient or inaccurate trajectory information. These 2-point shots are not included in our prediction model and thus 2-point FG\% is only partially Rao-Blackwellized. Interestingly, Rao-Blackwellizing FT\% also resulted in only a minor improvement in prediction. This is not due to lack of trajectory information as most free throws are included in our shot-make model, but may be because free throws more closely follow a Bernoulli distribution than either 2-point or 3-point shots. Free throws are certainly more homogeous than other shot attempts as they are not affected by contextual factors like changing shot distance or defender pressure. There has been some research showing serial correlation between free throws (Arkes 2010). Though even when shown this effect is considerably smaller than the effects contextual factors have on field-goal shot-make probabilities. The closer that a player's free throw attempts follow a Bernoulli distribution, the less potential there is to decrease the mean-squared error of the raw estimator of FT\% through Rao-Blackwellization. If a player's free throw attempts perfectly follow a Bernoulli distribution the number of makes and misses becomes a sufficient statistic for FT\% and Rao-Blackwellizing would give no improvement in prediction accuracy.

\vspace{1cm}
\begin{table}[ht]
\caption{Mean Absolute Prediction Errors of FG\% Estimators}
\centering
\begin{tabular}{cccccc}
  \hline \hline
 & Raw & Grand Mean & RB & Shrunk Raw & Shrunk RB \Tstrut\Bstrut\\
  \hline
  3-point shots & 0.0790 & 0.0620 & 0.0590 & 0.0689 & 0.0572 \Tstrut\\ [1ex]
  Free throws & 0.0809 & 0.0834 & 0.0713 & 0.0702 &\hspace{-0.28cm} 0.0691 \\ [1ex]
  2-point shots & 0.0549 & 0.0486 & 0.0502 & 0.0440 & 0.0428 \Bstrut\\ [0.5ex]
 \hline 
 \vspace{0.2cm}
  True-Shooting & 0.0467 & 0.0436 & 0.0408 & 0.0417 & 0.0379 \Tstrut\Bstrut\\ [-0.3ex]
   \hline
\end{tabular}
\begin{flushleft}
Estimators are for FG\% in the first half of the 2015-16 NBA season, errors based on prediction of FG\% in the second half of 2015-16. The raw estimator uses make/miss data, while the Rao-Blackwell (RB) estimator uses predicted shot-make probabilities. 
\end{flushleft}
\end{table}
\vspace{0.1cm}

\subsection{Example of an Improvement in Inferring Player FG\%}

We now present an example of when evaluating a player using $\hat{\theta}_{RB}$ instead of $\hat{\theta}$ may change the interpretation of that player's shooting and prediction of their future FG\%. After signing with Miami Lebron James improved his 3-point shooting ability drastically, shooting 36.5\% from three during the 2010-11 to 2014-15 seasons compared to just 32.9\% in his first 7 seasons in Cleveland (Paine 2016). However, during the 2015-16 season Lebron shot just 30.9\% from three. Was there a real difference in his 3-point shooting ability during this season compared to the previous 5? If we attempt to answer this question using raw FG\%, we can estimate a 90\% confidence interval via a normal approximation of $(0.264, 0.354)$. Thus with 90\% confidence we can say there was a real difference between Lebron's 3-point shooting during 2015-16 compared to the previous 5 years. More traditional advanced metrics also fail to explain James's dip in 3-point FG\%. Compared to the 2014-15 season (where James shot 35.4\% from three), in 2015-16 he shot from more favorable 3-point zones, shot fewer threes late in the shot clock, more of his threes came from assists, and fewer threes came against "tight" defensive pressure as classified by the SportVu tracking data (Paine 2016). All these indicators suggest that James's 3-point shooting should have improved in 2015-16, yet he shot his poorest percentage since his rookie year. Based on these statistics, one may have concluded that there was a real decrease in 3-point shooting skill during the 2015-16 season, and we may have predicted that this poor shooting would continue in upcoming seasons. However, if we instead use RB-FG\% as an estimator of 3P\%, we estimate his 3-point percentage during 2015-16 to be 34.7\%, with a 90\% confidence interval of $(0.321,0.374)$. Therefore, according to his RB-FG\% Lebron did not have an appreciable decline in 3-point shooting ability, and we would predict that his FG\% should revert back to somewhere around his average over the previous 5 years. As we've seen, this has indeed been the case as his 3P\% returned to 36.3\% and 36.7\% during the 2016-17 and 2017-18 seasons, respectively. 
\par

\section{Discussion and Conclusion}

In this paper we were able to construct an improved estimator for FG\% based on shot-make probabilities calculated from shot trajectories. Via the Rao-Blackwell theorem, we demonstrated that if we model shots according to a Beta-Bernoulli distribution, rather than a Bernoulli, aggregating shot-make probabilities for individual players is a more accurate estimator for future shooting than raw FG\%. Shot trajectory data has been shown to improve estimation of FG\% in other contexts. Marty (2018) demonstrates, using precise shot data captured by Noahlytics during practice shooting sessions, that raw shooting percentage augmented with 9 spatial rim patterns is a better estimate of shooting skill than raw FG\%. We are able to extend this idea to live-games, and show that shot features measured using the less precise optical tracking data can still provide improvement in FG\% prediction and estimation. Our method differs in that we create a new shooting statistic, one based on shot-make probabilities only, rather than use raw FG\% augmented with spatial features. Comparing the estimation ability of $\hat{\theta}_{RB}$ and Marty's raw FG\% augmented with spatial features is not explored in this paper, but both methods show distinct improvements when performing estimation on low sample sizes.  \par

Another way to quantify the quality of our Rao-Blackwellized metrics is to measure how well they are able to discriminate between players. We can accomplish this by comparing the discrimination meta-metric for Rao-Blackwellized and raw shooting metrics (Franks et al. 2016). This meta-metric quantifies the fraction of variance between players that is due to differences in true shooting skill. Table 3 shows that RB-3P\% and RB-TS\% are both more discriminative metrics than their raw counterparts. Franks et al. (2016) also define the meta-metric stability: the fraction of total variance in a metric that is due to true changes in player skill over time, rather than chance variability. We did not calculate this meta-metric as we do not have enough seasons of trajectory data to obtain accurate estimates.  \par

\begin{table}[b]
\caption{Discrimination Values for Raw and Rao-Blackwellized Shooting Metrics}
\centering
\begin{tabular}{c | c c | c c }
  \hline \hline
  & Raw 3P\% & RB-3P\% & Raw TS\% & RB-TS\% \Tstrut\Bstrut\\
  \hline
  Discrimination & 0.432 & 0.548 & 0.713 & 0.804 \Tstrut\\ [1ex]
   \hline
\end{tabular}
\begin{flushleft}
Estimates are based on Discrimination metrics for the 2014-15 season. The RB metrics are shrunk as defined in Section 4.1.
\end{flushleft}
\end{table}
\vspace{0.1cm}

There have been many other models that use game-specific context variables like defender distance and shot location to try and estimate the probability that shots will go in (Cen et al. 2015, Chang et al. 2014). These models are also Rao-Blackwellizing FG\%, as they are assuming shot probabilities vary for each shot (see Section 2). However, $\hat{\theta}_{RB}$ should still improve on these models because our estimated shot factors are sufficient for all in-game contextual variables that contribute to shot-make probabilities. Including the location of the shot or the nearest defender distance should not change the probability a shot will go in given its depth, left-right accuracy, and entry angle at the basket. We are able to classify shots correctly 79.6\% of the time using predicted make probabilities based on trajectory information, higher than the 61\% classfication rate we found using nearest defender distance and shot location as predictors of raw FG\%, and also higher than those found in more complex contextual models (Cen et al. 2015, Chang et al. 2014). Additionally, adding shot-distance and nearest-defender distance as dimensions to RB-FG\% did not improve classification. \par

Because RB-FG\% allows us to more accurately estimate true FG\% with smaller sample sizes, we should be able to more accurately predict how contextual shooting variables like defender distance impact a player's shooting. Unfortunately, it is difficult to compare coefficients for contextual variables when fitting predicted probabilities compared to a binary shot response (make/miss) because we are estimating coefficients using different loss functions. Therefore, when we try to compare these coefficient estimates to a "true" value, for example how defender-distance affects FG\% for a player over the entire season, we are comparing two estimated coefficients to a "true" coefficient value which is also estimated using a binary shot response. Even if the coefficient for defender distance estimated using $\hat{\theta}_{RB}$ as a response is a better indicator of how a player responds to defensive pressure, it is difficult to compare this to any standard value for that player. \par

Although all NBA teams almost exclusively use raw FG\% and its aggregate statistics to evaluate player shooting, many teams use shot trajectory characteristics to evaluate and coach player shooting in practice. The Noah Shooting System is used by a number of teams to analyze player shooting and to improve shot trajectories during practice shooting sessions. Analysis of trajectories in games, however, is not typically done due to the noisiness of the location data in the SportVu database. This paper provides a method to utilize in-game shot trajectories provided by the optical tracking data to better evaluate and predict player shooting. \par

\section*{References}
\singlespacing
\setlength{\parskip}{1em}
\begin{hangparas}{.5in}{1}

Aesrk, J. 2010. "Revisiting the Hot Hand Theory with Free Throw Data in a Multivariate Framework." \textit{Journal of Quantitative Analysis in Sports} 6(1): Retrieved 12 Jun. 2018, from doi:10.2202/1559-0410.1198.

Blackport, D. 2014. "How Long Does it Take for Three Point Shooting to Stabilize?" https://fansided.com/-2014/08/29/long-take-three-point-shooting-stabilize/. Accessed November 11th, 2017. 

Brown, L.D. 2008. "In-Season Prediction of Batting Averages: A Field Test of Empirical Bayes and Bayes Methodologies." \textit{The Annals of Applied Statistics} 2:113-152. 

Casella, G. 1985. "An Introduction to Empirical Bayes Data Analysis." \textit{The American Statistician} 39:83-87.

Chang, Y.H., R. Maheswaran, J. Su, S. Kwok, T. Levy, A. Wexler, and K. Squire. 2014. "Quantifying Shot Quality in the NBA." \textit{Proceedings of the 2014 MIT Sloan Sports Analytics Conference}.

Cen, R., H. Chase, C. Pena-Lobel, and D. Silberwasser. 2015. "NBA Shot Prediction and Analysis." https://hwchase17.github.io/sportvu/. Accessed November 11th, 2017.

Efron, B., and C. Morris. 1977. "Stein's paradox in statistics." \textit{Scientific American} 236:119-127.  

Franks, A., A. D'Amour, D. Cervone, and L. Bornn. 2016. "Meta-Analytics: Tools for Understanding the Statistical Properties of Sports Metrics." \textit{Journal of Quantitative Analysis in Sports} 12:151-165.

Fieller, E. C., Hartley, H. O., and E.S. Pearson. 1957. "Tests for rank correlation coefficients." \textit{Biometrika} 44:470-481.

Griffiths, D.A. 1973.  "Maximum Likelihood Estimation for the Beta-Binomial Distribution and an Application to the Household Distribution of the Total Number of Cases of a Disease." \textit{Biometrics} 637-648.

Kubatko, J., D. Oliver, K. Pelton, and D.T. Rosenbaum. 2007. "A Starting Point for Analyzing Basketball Statistics." \textit{Journal of Quantitative Analysis in Sports} 3(3): Retrieved 12 Jun. 2018, from doi:10.2202/1559-0410.1070.

Marty, R. 2018. "High-resolution Shot Capture Reveals Systematic Biases and an Improved Method for Shooter Evalutation." \textit{Proceedings of the 2018 MIT Sloan Sports Analytics Conference}.

Marty, R. and S. Lucey. 2017. "A Data-Driven Method for Understanding and Increasing 3-Point Shooting Percentage." \textit{Proceedings of the 2017 MIT Sloan Sports Analytics Conference}.

Paine, N. 2016. "LeBron's 3-Point Shot Has Abandoned Him." https://fivethirtyeight.com/features/lebrons-3-point-shot-has-abandoned-him/. Accessed January 3rd, 2018.

Piette, J., A. Sathyanarayan, and K. Zhang. 2010. "Scoring and Shooting Abilities of NBA Players." \textit{Journal of Quantitative Analysis in Sports} 6(1): Retrieved 12 Jun. 2018, from doi:10.2202/1559-0410.1194.

Skellam, J.G. 1948. "A Probability Distribution Derived from the Binomial Distribution by Regarding the Probability of Success as Variable Between the Sets of Trials." \textit{Journal of the Royal Statistical Society. Series B (Methodological)} 10:257-261.

Young, S. 2016. "The NBA's 3-point Revolution" https://bballbreakdown.com/2016/12/16/the-nba-3-point-revolution/. Accessed December 14th, 2017.  

\end{hangparas}
\end{document}